\documentclass[aps,prd,nofootinbib,superscriptaddress,floatfix]{revtex4}
\usepackage{graphicx}
\usepackage{bm}
\usepackage{braket}
\usepackage{bbold}
\usepackage{amssymb}
\usepackage{amsmath,latexsym}
\usepackage{subfigure}
\usepackage{slashed}
\usepackage[T1]{fontenc}
\usepackage{multirow,array}
\usepackage{mathtools}
\usepackage{mathrsfs}
\usepackage{dsfont}
\usepackage[colorlinks=false,linktocpage=true]{hyperref}
\usepackage{hyperref}
\usepackage[utf8]{inputenc}
\usepackage[export]{adjustbox}
\usepackage{float}
\usepackage[makeroom]{cancel}
\usepackage{wrapfig}
\usepackage{ulem}
\usepackage{color}
\newcommand{\DD}{\mathcal D}
\newcommand{\DI}{\mathcal D^{\mathrm{IMF}}}
\newcommand{\DDbar}{\mathcal{D}}

\newcommand{\lap}{\nabla_\perp^2}
\newcommand{\bvec}{\bm b}
\newcommand{\Dvec}{\bm\Delta}

\newcommand{\be}{\begin{equation}}
\newcommand{\ee}{\end{equation}}
\newcommand{\ba}{\begin{eqnarray}}
\newcommand{\ea}{\end{eqnarray}}

\newcommand{\uncomment}[1]{{ \red{\bf[text removed]} }}

\begin{document}
\title{\boldmath
An invertible map between 3D Breit-frame mechanical distributions and 2D infinite-momentum-frame mechanical densities in spin-1 hadrons}
\author{Kemal Tezgin}
\email{kemaltezgin@gmail.com}
\affiliation{Department of Physics, University of Virginia, Charlottesville, VA 22904, USA}
\affiliation{Department of Physics, Virginia Tech, Blacksburg, VA 24061, USA}

\date{09/17/2026}
\begin{abstract}
\noindent
We construct an invertible map between the complete set of 3D Breit-frame pressure and shear force distributions and the constrained space of 2D infinite-momentum-frame pressure and shear force densities in coordinate space for spin-1 hadrons. The correspondence between the two sets of distributions in different frames is established by combining the geometric forward and inverse Abel transformations with recoil operators after separating the Wigner rotation induced mixing effects on the infinite-momentum-frame side. Unlike in the spin-1/2 case, these mixings complicate the coordinate space relations for spin-1 hadrons by introducing additional multipole structures, but the constraints among the form factors propagate into constraints on mechanical densities, so that these additional multipoles do not introduce new mechanical information.
\end{abstract}
\maketitle


\maketitle

\section{Introduction}\label{Introduction}
  
Hadronic matrix elements of the QCD energy-momentum tensor (EMT) are parametrized by EMT form factors \cite{Kobzarev:1962wt, Pagels:1966zza, Ji:1996ek} (often also referred to as gravitational form factors, GFFs, because EMT is the source that couples to gravity). Similar to electromagnetic form factors, which encode the charge and magnetization distributions after Fourier transformation, EMT form factors encode some other spatial distributions within hadrons, namely the mass, spin, pressure, and shear force distributions \cite{Polyakov:2002yz, Polyakov:2018zvc}. However, these distributions manifest themselves in particular frames. To obtain the three-dimensional spatial distributions, one needs to employ the Breit frame, in which only the spatial momentum is transferred to the system, while keeping the average momentum of the hadron at zero. However, for hadrons with a comparable Compton wavelength to their characteristic spatial size, this picture is blurred by relativistic recoil corrections and localization effects \cite{Miller:2018ybm, Jaffe:2020ebz, Freese:2021czn}. One can rather adopt a phase-space approach to assign quasi-probabilistic interpretations to Breit-frame distributions \cite{Lorce:2018egm, Lorce:2020onh}. On the other hand, with the elastic frame, one has the freedom to vary the average momentum component along the $z$-direction, $P_z$, while keeping the momentum transfer along the $z$-direction vanishing. In such frames, one could then interpolate between the two-dimensional Breit frame (BF), $P_z=0$, and the two-dimensional infinite momentum frame (IMF), equivalently on the light-front, $P_z\to\infty$, in a controlled way \cite{Lorce:2017wkb, Lorce:2018egm, Lorce:2020onh}.   

It is therefore convenient to find an invertible map that relates a distribution obtained in one frame to another in another frame. For spin-1/2 hadrons, the one-to-one relation between 3D BF distributions and 2D IMF densities is essentially simple, as there is only one D-term form factor that appears in the parametrization of the EMT matrix elements. A single Abel transformation can transform a 3D BF mechanical distribution to a 2D IMF mechanical density \cite{Panteleeva:2021iip}. Conversely, using the inverse Abel transformation, one can obtain a 3D BF mechanical distribution from a 2D IMF mechanical density. 
 
For a spin-1 target, however, their connection is not visible by a simple Abel transformation as tensor polarization introduces additional mechanical multipole form factors. In the BF, the decomposition of the EMT matrix element contains three multipole form factors, whereas the IMF decomposition yields five multipole form factors due to the Wigner spin rotations under Lorentz boosts \cite{Kim:2022wkc}. Moreover, as in the spin-0 case \cite{Freese:2021mzg}, recoil corrections are also apparent in the spin-1 mechanical distributions and further spoil the simple relations between these two frames. 

We argue that such boost effects do not introduce additional mechanical information. The five IMF multipoles are subject to two constraints and, hence, can be described by three functions. At the density level, this means that, after an appropriate recombination of 2D IMF mechanical densities, one can reproduce the 2D BF pressure and shear forces. After restoring the recoil effects, the inverse Abel transformation yields the 3D BF distributions. Conversely, starting with 3D BF mechanical distributions, the dipole-induced IMF density can be obtained from the 2D BF distributions (which are obtained by Abel transformation of 3D BF distributions) by first removing the recoil effects and then solving an inhomogeneous Helmholtz equation. Having the dipole-induced IMF density is then sufficient to obtain the remaining IMF densities. In this sense, the mapping from 3D BF mechanical distributions to 2D IMF densities is invertible. 

There is a growing literature on EMT form factors, their associated 3D BF distributions, and 2D IMF densities of spin-1 hadrons, with different aspects of the subject studied, for example, in \cite{Kim:2022wkc, Abidin:2008ku, Taneja:2011sy, Cosyn:2019aio, Polyakov:2019lbq, Freese:2019bhb, Sun:2020wfo, Epelbaum:2021ahi, Freese:2022yur, He:2023ogg, He:2024vzz, GarciaMartin-Caro:2023klo, GarciaMartin-Caro:2023toa, Panteleeva:2024abz, Cosyn:2026gyy}. In particular, in \cite{Freese:2022yur} it also studied the 2D spatial densities of pressure and shear forces in spin-1 hadrons within the light-front formalism. Lattice QCD results of gluon GFFs of various hadrons, including the spin-1 $\rho$ meson, were studied in \cite{Pefkou:2021fni}. In addition, the Abel transformation in relation to the nucleon has been studied in \cite{Freese:2021mzg, Kim:2021jjf, Kim:2022syw, Choudhary:2022den}, while in relation to the electromagnetic multipole structure of a spin-1 target has been studied in \cite{Kim:2022bia}. For a general framework for covariant multipole decomposition of higher spin matrix elements, we refer to \cite{Cosyn:2025gmp}.

\section{Breit-frame decomposition of the spin-1 EMT matrix elements}

For an incoming spin-1 hadron with four-momentum $p$ and the outgoing one with $p'$, let us denote the average four-momentum of the system as $P=(p+p')/2$, the four-momentum transfer as $\Delta=p'-p$, and the invariant momentum transfer as $t=\Delta^2$. Then, the total conserved EMT matrix element can be parametrized by six EMT form factors as \cite{Abidin:2008ku, Taneja:2011sy, Cosyn:2019aio, Polyakov:2019lbq}
\begin{align}
    \langle p',\lambda'|\hat T^{\mu\nu}(0)|p,\lambda\rangle
    = & 2P^\mu P^\nu \left[-\varepsilon'^{*}\!\cdot\varepsilon\,A_0(t)
    +\frac{(\varepsilon'^{*}\!\cdot P)(\varepsilon\cdot P)}{m^2}A_1(t)\right] \nonumber\\
    & +2\Bigl[P^\mu\bigl(\varepsilon'^{*\nu}\,\varepsilon\cdot P
    +\varepsilon^\nu\,\varepsilon'^{*}\!\cdot P\bigr) + P^\nu\bigl(\varepsilon'^{*\mu}\,\varepsilon\cdot P
    +\varepsilon^\mu\,\varepsilon'^{*}\!\cdot P\bigr)\Bigr]J(t) \nonumber\\
    & +\frac{1}{2}\bigl(\Delta^\mu\Delta^\nu-\eta^{\mu\nu}\Delta^2\bigr)
    \left[\varepsilon'^{*}\!\cdot\varepsilon\,D_0(t)+\frac{(\varepsilon'^{*}\!\cdot P)(\varepsilon\cdot P)}{m^2}D_1(t)\right] \nonumber\\
    & +\Bigl[\frac{1}{2}\bigl(\varepsilon^\mu\varepsilon'^{*\nu}+\varepsilon'^{*\mu}\varepsilon^\nu\bigr)\Delta^2
    -\bigl(\varepsilon'^{*\mu}\Delta^\nu+\varepsilon'^{*\nu}\Delta^\mu\bigr)\varepsilon\cdot P\nonumber\\
    &\hspace{2.5cm}+\bigl(\varepsilon^\mu\Delta^\nu+\varepsilon^\nu\Delta^\mu\bigr)\varepsilon'^{*}\!\cdot P
    -4\eta^{\mu\nu}(\varepsilon'^{*}\!\cdot P)(\varepsilon\cdot P)\Bigr]E(t) .
\end{align}
where we denote the initial and final polarization vectors as $\varepsilon^{\mu}=\varepsilon^\mu(p,\lambda)$ and $\varepsilon'^\mu=\varepsilon'^\mu(p',\lambda')$, respectively. We also use the covariant normalization $\langle p',\lambda'|p,\lambda\rangle = 2 p^0 (2\pi)^3 \delta_{\lambda'\lambda} \delta^{(3)}(\bm{p'}-\bm{p})$ of one-particle states. There are three additional form factors that appear at the partonic level but vanish in the total EMT matrix decomposition, see Refs.~\cite{Cosyn:2019aio, Polyakov:2019lbq} for details. Throughout this work, we will focus on the $ij$-components of the EMT, often referred to as $D$-term form factors. Unlike their spin-0 and spin-1/2 counterparts, where there appears only one $D$ form factor in the parametrization of the total EMT, spin-1 hadrons have three $D$-term form factors, denoted here, following the notations in \cite{Polyakov:2019lbq}, as $D_0(t)$, $D_1(t)$, and $E(t)$. These form factors encode the distributions of pressure and shear forces within spin-1 hadrons in certain frames. 

Since form factors solely depend on the four-momentum exchange, the three-dimensional spatial information contained in these form factors is often acquired in the Breit-frame (BF), where there is no energy exchange between the initial and final states, i.e. $\Delta^0 = 0$, and the average three-momenta of the collision is zero, i.e. $\bm{P} = \bm{0}$, so that $\bm{p'} = -\bm{p} = \bm{\Delta}/2$. As presented in \cite{Polyakov:2019lbq} (see also \cite{Kim:2022wkc}), the $ij$-component of the EMT matrix element in the BF, evaluated between canonical spin states, can be expressed in a multipole basis as 
\begin{align}
    \langle p',\lambda'|\hat T^{ij}(0)|p,\lambda\rangle &= 2m^2\tau \left(Y^{ij}_2(\Omega_\Delta) 
    - \frac{2}{3}\delta^{ij}\right)\mathcal D_0(t)\,\delta_{\lambda'\lambda}\nonumber \\
    &\quad + 8\tau^2m^2 \left(Y^{ij}_2(\Omega_\Delta) -\frac{2}{3}\delta^{ij}\right)
    Y^{kl}_2(\Omega_\Delta)\hat Q^{kl}_{\lambda'\lambda} \mathcal D_3(t)\nonumber\\
    &\quad+ 4m^2\tau \left[Y^{jk}_2(\Omega_\Delta)\hat Q^{ik}_{\lambda'\lambda} +Y^{ik}_2(\Omega_\Delta)\hat Q^{jk}_{\lambda'\lambda}
    - \frac{1}{3}\hat Q^{ij}_{\lambda'\lambda} - \delta^{ij}Y^{kl}_2(\Omega_\Delta)\hat Q^{kl}_{\lambda'\lambda} \right]\mathcal D_2(t) \, .
\end{align}
where we denote the quadrupole operator of spin-1 particles in terms of spin operators $\hat{S}^i$ as 
\begin{equation}
    \hat{Q}^{ij} = \frac{1}{2}\left(\hat{S}^i\hat{S}^j + \hat{S}^j\hat{S}^i -\frac{2}{3} S (S+1)\delta^{ij} \right) \, ,
\end{equation}
and use the notation $Y^{ij}_2(\Omega_\Delta) = \frac{\Delta^i\Delta^j}{\Delta^2} - \frac{1}{3} \delta^{ij}$ for the rank-2 irreducible Cartesian tensor. Each BF mechanical multipole form factor represents a certain combination of spin-1 $D$-term form factors and is given by \cite{Polyakov:2019lbq}
\begin{align}
  \DD_0(t) &= -D_0(t)+\frac{4}{3}E(t)-\frac{\tau}{3}\left[2D_0(t)-2E(t)+D_1(t)\right]-\frac{\tau^2}{3}D_1(t),\label{D0-BF}\\
  \DD_2(t) &= -E(t),\label{D2-BF}\\
  \DD_3(t) &= \frac{1}{4}\left[2D_0(t)-2E(t)+D_1(t)\right]+\frac{\tau}{4}D_1(t), \label{D3-BF}
\end{align}
with $\tau=-t/(4m^2)$. Then, one can obtain the 3D spatial distributions by taking the inverse Fourier transform of these multipoles with respect to the spatial momentum exchange as 
\begin{equation}
  \widetilde{\DD}_n(r) = \int\frac{d^3\Dvec}{(2\pi)^3} e^{-i\Dvec\cdot\bm r}\frac{m}{E} \DD_n(t), \qquad n=0,2,3.
\end{equation}
where, the energy term, $E$, introduces an additional $\Dvec$ dependence inside the integral through the relation $E=\sqrt{m^2-t/4}$. We call $\widetilde{\DD}_n(r)$ the 3D BF potential, as it can be used as the source for decomposing each type into a trace and a traceless part, yielding the pressure and shear-forces inside the spin-1 particle, respectively, under the operations  
\begin{align}
  p_n(r) &= \frac{1}{6m}\frac1{r^2}\frac{d}{dr} r^2\frac{d}{dr}\widetilde{\DD}_n(r), \label{pressure_3d}\\
  s_n(r) &= -\frac{1}{4m}r\frac{d}{dr}\frac1r\frac{d}{dr}\widetilde{\DD}_n(r). \label{shear_3d}
\end{align}
So, unlike in spin-0 and spin-1/2 cases, where only one associated pressure and shear force distribution appears, in spin-1 there are three associated distributions in the Breit frame. It is also well-known in the literature that, for each $n$, due to the total EMT conservation, pressure and shear forces satisfy the equilibrium conditions \cite{Polyakov:2018zvc, Polyakov:2019lbq}
\begin{equation}
  p'_n(r)+\frac23s'_n(r)+\frac2r s_n(r)=0, \qquad n=0,2,3.
\end{equation}

\section{Infinite-momentum-frame decomposition of the spin-1 EMT matrix elements}

As noted earlier, the 3D distributions have inherent problems when referred to as densities. Especially because the Compton wavelength of the proton, for instance, is comparable to its intrinsic size, probing sizes shorter than its wavelength demands relativistic recoil corrections. Hence, the interpretation of these distributions has attracted multiple criticisms for treating them as densities \cite{Miller:2018ybm, Jaffe:2020ebz, Freese:2021czn}. Localized wave packet analyses also assert that the Breit frame distributions correspond to a static approximation \cite{Epelbaum:2022fjc, Panteleeva:2022uii, Panteleeva:2023evj}. On the other hand, by adopting a phase-space approach, it is possible that one can assign a quasi-probabilistic meaning to those 3D BF distributions \cite{Lorce:2018egm, Lorce:2020onh}. It is also worth noting that such criticisms for 3D BF distributions become less prominent for heavy nuclei or in the large-$N_c$ limit as they acquire a 3D density interpretation \cite{Goeke:2007fp, Polyakov:2018zvc, Neubelt:2019sou, Lorce:2022cle}. For more details on the use of Breit frame distributions and their interpretation, we refer the reader to the recent discussion in \cite{Lorce:2025oot} and the references therein.   

One way to overcome such recoil effects from the beginning is to define 2D transverse distributions in the infinite-momentum frame (IMF). Such a frame corresponds to, in addition to having no energy exchange, taking the average momentum in the $z$-direction to infinity, i.e., $P_z \to \infty$, so that one needs to restrict themselves to the zero momentum exchange along the $z$-direction, $\Delta_z = 0$, to preserve the on-shellness condition $P\cdot\Delta = 0$. For spin-1 particles, the mechanical structures in different frames are analyzed in detail in Ref.~\cite{Kim:2022wkc}. By using the canonical spin states in the elastic frame and taking the limit $P_z \to \infty$, one obtains the following IMF multipole expansion of the matrix element in the transverse $ij$-components \cite{Kim:2022wkc}
\begin{align}
  \langle p',\lambda'|\hat T^{ij}(0)|p,\lambda\rangle &= 2m^2\tau 
  \left[ \left(\frac13 \DI_2(t) -\frac12 \DI_{(0,1)}(t)\right) \delta_{\sigma'\sigma}
  + \left(-\frac23 \DI_2(t) -\frac12 \DI_{(0,0)}(t)\right)\delta_{\lambda'3}\delta_{\lambda3}\right]\delta^{ij}\nonumber\\
  &\quad \quad +2m^2\tau X_2^{ij}(\theta_{\Delta_\perp})
  \left[\delta_{\sigma'\sigma} \DI_{(0,1)}(t)+\delta_{\lambda'3}\delta_{\lambda3} \DI_{(0,0)}(t)\right]\nonumber\\
  &\quad \quad + 4m^2\tau
  \left[\hat Q^{ik}_{\lambda'\lambda} X_2^{jk}(\theta_{\Delta_\perp}) +\hat Q^{jk}_{\lambda'\lambda} X_2^{ik}(\theta_{\Delta_\perp}) - \hat Q^{lm}_{\lambda'\lambda} X_2^{lm}(\theta_{\Delta_\perp})\delta^{ij}\right] \DI_2(t)\nonumber\\
  &\quad \quad +8m^2\tau^{3/2}i\epsilon^{lm3}\hat S^l_{\lambda'\lambda} X_1^m (\theta_{\Delta_\perp}) \left(X_2^{ij}(\theta_{\Delta_\perp})-\frac12\delta^{ij}\right) \DI_1(t) \nonumber\\
  &\quad \quad + 8m^2\tau^2\hat Q^{lm}_{\lambda'\lambda} \left(X_2^{lm}(\theta_{\Delta_\perp})+\frac12\delta^{lm}\right)
  \left(X_2^{ij}(\theta_{\Delta_\perp})-\frac12\delta^{ij}\right) \DI_3(t) \, . \label{EMT_IMF}
\end{align}
Here, $i,j,k,l,m = 1,2$ denote the transverse indices, and $\lambda, \lambda'=1,2,3$ denote the spin states, whereas $\sigma, \sigma' = 1,2$ are restricted to the transverse polarizations. We also use the two-dimensional irreducible tensors $X_1^{i}(\theta_{\Delta_\perp}) = \frac{\Delta_\perp^i}{\Delta_\perp}$ and $X_2^{ij}(\theta_{\Delta_\perp}) = \frac{\Delta_\perp^i\Delta_\perp^j}{\Delta_\perp^2} - \frac{1}{2} \delta^{ij}$. So, unlike the BF case, in the IMF one obtains five multipole form factors due to Wigner spin rotations under Lorentz boosts. But those five IMF multipole form factors in Eq.~(\ref{EMT_IMF}) can be expressed in terms of the BF multipole form factors as \cite{Kim:2022wkc}   
\begin{align}
  \DI_{(0,0)} &= \DD_0+\frac{4\tau}{3}G_W, \qquad \DI_{(0,1)} = \DD_0+\frac{\tau}{3}G_W, \nonumber \\
  \DI_1 &= \frac14G_W,  \qquad \DI_2 = \DD_2, \qquad \DI_3 = \DD_3-\frac14G_W, \label{IMF_multipoles}
\end{align}
where
\begin{equation}
    G_W = -\frac{2(3\DD_0+\DD_2-2\tau\DD_3)}{3(1+\tau)} \, . \label{GW}
\end{equation}
Here, the effects of Wigner spin rotation can be parametrized by $G_W$, a specific combination of BF multipoles. In particular, among all IMF multipoles, $\DI_1$ is purely a relativistically induced IMF dipole. It purely arises from the Wigner rotation of canonical spin states under boost from BF to IMF. 

We observe that the five IMF spin-1 mechanical multipoles contain only three independent functions, as they obey the two constraints
\begin{align}
  \DI_{(0,0)}-\DI_{(0,1)} - 4\tau\DI_1 &= 0 \, , \label{2D_FFconstraints_1} \\
  3\DI_{(0,1)}+\DI_2-2\tau\DI_3+6\DI_1 &=0 \, . \label{2D_FFconstraints_2}
\end{align}
Conversely, by having any set of IMF mechanical multipoles satisfying these constraints, one can reconstruct the three BF multipole form factors through 
\begin{equation}
  \DD_2 = \DI_2,  \qquad \DD_3 = \DI_3+\DI_1,  \qquad \DD_0 = \DI_{(0,1)}-\frac{4\tau}{3}\DI_1 \, . 
\end{equation}

Therefore, even though there seems to be five independent IMF multipoles, there is an invertible map between the three BF multipoles and the five IMF ones at the form factor level, provided that the constraints in Eqs.~(\ref{2D_FFconstraints_1}--\ref{2D_FFconstraints_2}) are satisfied. 

\section{Relations between 2D mechanical potentials}

In the previous section, we showed the relations between the two sets of multipole form factors obtained in BF and IMF. Next, we want to explicate how the Abel transformation of the 3D BF potential is encoded in the 2D IMF potential, as this will set the bridge between the mechanical forces in these different frames. As shown in \cite{Panteleeva:2021iip}, there is a one-to-one correspondence between 3D BF distributions and 2D IMF densities via Abel tomography for mechanical forces within spin-1/2 hadrons. This can be seen from the appearance of a single $D$ form factor in the parametrization of the nucleon matrix element and the absence of the recoil term $m/E$. For spin-1 hadrons, the situation is different because the IMF multipoles contain Wigner spin rotation-induced mixing and the recoil term in the BF Fourier transforms. Therefore, there are two consecutive operations to be performed when moving from 3D BF to 2D IMF; first, after applying the Abel transformation, one needs to convert recoil-included 2D BF terms to the recoil-reduced ones, and then one has to construct the corresponding mixings. Similarly, when moving from 2D IMF to 3D BF, one first needs to undo these mixings on the IMF side, then convert them to 2D recoil-included BF ones before applying the inverse Abel transformation. 

Now, let us move to the position space and denote the two-dimensional Fourier transform of $\DI_\alpha(t)$ by
\begin{equation}
    \widetilde{\DD}^{\mathrm{IMF}}_\alpha(b)=\int\frac{d^2\bm\Delta_\perp}{(2\pi)^2}\, e^{-i\bm\Delta_\perp\cdot\bvec}\,\DI_\alpha(t),\quad \text{with} \quad t=-\bm\Delta_\perp^2 .
\end{equation}
and call it the 2D IMF potential. This potential is obtained by following the normalization in Ref.~\cite{Kim:2022wkc}, which scales the stress tensor in the IMF limit by $P^0/m$ to have a finite term at the integrand level. On the other hand, we define $\widetilde{\DDbar}_n(b)$ as the Abel transform of the 3D BF potential
\begin{equation}
  \widetilde{\DDbar}_n(b) = \int_{-\infty}^{\infty}dz\, \widetilde{\DD}_n\!\left(\sqrt{b^2+z^2}\right)
  = 2\int_b^\infty dr\, \frac{r\,\widetilde{\DD}_n(r)}{\sqrt{r^2-b^2}} \,,
\end{equation}
which can be written equivalently as the two-dimensional Fourier transform of the same BF multipole at $\Delta_z=0$
\begin{equation}
     \widetilde{\DDbar}_n(b) = \int\frac{d^2\Dvec_\perp}{(2\pi)^2} \, e^{-i\Dvec_\perp\cdot\bm b} \, \frac{m}{E} \, \DD_n(-\Delta_\perp^2) \, .
\end{equation}
For that, $\widetilde{\DDbar}_n(b)$ is the 2D BF potential (we use the parameter $b$ to reflect its two-dimensional character). In making connections between 2D BF distributions and 2D IMF ones, we also introduce a reduced 2D BF potential without the recoil term $m/E$  
\begin{equation}
     \widetilde{\DDbar}^{\text{red}}_n(b) = \int\frac{d^2\Dvec_\perp}{(2\pi)^2} e^{-i\Dvec_\perp\cdot\bm b}\DD_n(-\Delta_\perp^2) 
\end{equation}
and they are related to each other by
\begin{equation}
     \widetilde{\DDbar}_n(b) = \bigg(1- \frac{\lap}{4m^2}\bigg)^{-1/2} \widetilde{\DDbar}^{\text{red}}_n(b) \, .
\end{equation}
Since in the transverse space multiplication by $\tau$ acts as $-\frac{\lap}{4m^2}$, the reduced 2D BF potentials are related to the 2D IMF potentials as follows
\begin{align}
  \widetilde{\DDbar}^{\text{red}}_2(b) &= \widetilde{\DD}^{\mathrm{IMF}}_2(b) \,, \label{2D_BF_potential_IMF_1} \\
  \widetilde{\DDbar}^{\text{red}}_3(b) &= \widetilde{\DD}^{\mathrm{IMF}}_3(b) + \widetilde{\DD}^{\mathrm{IMF}}_1(b) \,, \label{2D_BF_potential_IMF_2} \\
  \widetilde{\DDbar}^{\text{red}}_0(b) &= \widetilde{\DD}^{\mathrm{IMF}}_{(0,1)}(b) +\frac{1}{3m^2}\lap \widetilde{\DD}^{\mathrm{IMF}}_1(b) \,. \label{2D_BF_potential_IMF_3}
\end{align}
These relations identify the reduced 2D BF potentials with local combinations of the 2D IMF potentials. Thus, the 2D IMF potentials become Abel-invertible only after this recombination is performed and the recoil corrections are taken into account. From Eqs.~\ref{2D_FFconstraints_1}--\ref{2D_FFconstraints_2}, the two constraints on the IMF potential become
\begin{align}
  \widetilde{\DD}^{\mathrm{IMF}}_{(0,0)}(b) - \widetilde{\DD}^{\mathrm{IMF}}_{(0,1)}(b) + \frac{1}{m^2}\lap \widetilde{\DD}^{\mathrm{IMF}}_1(b)&= 0, \label{2D_Potential_Constraint_1}\\
  3\widetilde{\DD}^{\mathrm{IMF}}_{(0,1)}(b) + \widetilde{\DD}^{\mathrm{IMF}}_2(b) +\frac{1}{2m^2}\lap \widetilde{\DD}^{\mathrm{IMF}}_3(b)+ 6\widetilde{\DD}^{\mathrm{IMF}}_1(b) &=0. \label{2D_Potential_Constraint_2}
\end{align}


\section{Relations between 2D pressure and shear force densities}

In the previous section, we related the local combinations of IMF potentials to the reduced 2D BF potentials. In this section, we formulate the implications of the last section for the pressure and shear forces inside spin-1 hadrons. 

The 2D IMF pressure and shear force densities were introduced in \cite{Lorce:2018egm} for spin-1/2 particles. Similarly, they can be extended to spin-1 particles as was shown in \cite{Kim:2022wkc}. For a two-dimensional potential $D(b)$, pressure and shear forces can be expressed as
\begin{equation}
  p[D] = \frac{1}{8m}\left( \frac{d^2}{db^2}+\frac1b\frac{d}{db} \right) D, \qquad s[D] =-\frac{1}{4m}\left( \frac{d^2}{db^2}-\frac1b\frac{d}{db} \right) D .
\end{equation}
These operators define the pressure and shear forces associated with any two-dimensional mechanical potential in the present normalization. They can therefore be applied both to the IMF potentials $\widetilde{\DD}^{\mathrm{IMF}}_\alpha(b)$ and to the 2D BF potentials $\DDbar_n(b)$ and $\DDbar^{\text{red}}_n(b)$. To translate relations in Eqs.~(\ref{2D_BF_potential_IMF_1}--\ref{2D_BF_potential_IMF_3}) to the language of pressure and shear forces, we first note that  
\begin{align}
  p[\lap D] &= \lap p[D] \, , \\
  s[\lap D] &= \left(\lap-\frac{4}{b^2}\right)s[D] \, .
\end{align}
Therefore, one finds a local relation in obtaining 2D reduced BF pressure and shear force distributions from the IMF pressure and shear forces as
\begin{align}
  p^{\text{red}}_2(b) &= p^{\mathrm{IMF}}_2(b), & s^{\text{red}}_2(b) &= s^{\mathrm{IMF}}_2(b) \, , \label{2D_BF_from_IMF_forces_1} \\
  p^{\text{red}}_3(b) &= p^{\mathrm{IMF}}_3(b) +p^{\mathrm{IMF}}_1(b), & s^{\text{red}}_3(b) &= s^{\mathrm{IMF}}_3(b) +s^{\mathrm{IMF}}_1(b) \, , \label{2D_BF_from_IMF_forces_2} \\
  p^{\text{red}}_0(b) &= p^{\mathrm{IMF}}_{(0,1)}(b)+\frac{1}{3m^2}\lap p^{\mathrm{IMF}}_1(b), & s^{\text{red}}_0(b) &= s^{\mathrm{IMF}}_{(0,1)}(b) + \frac{1}{3m^2} \left(\lap-\frac{4}{b^2}\right)s^{\mathrm{IMF}}_1(b) \, \label{2D_BF_from_IMF_forces_3}.
\end{align}
Similarly, the constraints in Eqs.~(\ref{2D_Potential_Constraint_1}--\ref{2D_Potential_Constraint_2}) translate into the 2D IMF pressure and shear force relations as
\begin{align}
  p^{\mathrm{IMF}}_{(0,0)}(b) - p^{\mathrm{IMF}}_{(0,1)}(b) + \frac{1}{m^2}\lap p^{\mathrm{IMF}}_1(b) &= 0 \, , \label{2D_IMF_pressure_condition_1} \\ 
  s^{\mathrm{IMF}}_{(0,0)}(b) - s^{\mathrm{IMF}}_{(0,1)}(b) + \frac{1}{m^2}\left(\lap-\frac{4}{b^2}\right)s^{\mathrm{IMF}}_1(b) &= 0 \, ,  \label{2D_IMF_shear_condition_1} \\
  3p^{\mathrm{IMF}}_{(0,1)}(b) + p^{\mathrm{IMF}}_2(b) +\frac{1}{2m^2}\lap p^{\mathrm{IMF}}_3(b) +6p^{\mathrm{IMF}}_1(b) &=0 \, , \label{2D_IMF_pressure_condition_2} \\
  3s^{\mathrm{IMF}}_{(0,1)}(b) +s^{\mathrm{IMF}}_2(b) + \frac{1}{2m^2} \left(\lap-\frac{4}{b^2}\right)s^{\mathrm{IMF}}_3(b)+6s^{\mathrm{IMF}}_1(b) &=0 \, \label{2D_IMF_shear_condition_2} .
\end{align}
These equations explicitly show that, among the five IMF pressure and shear force densities, only three are linearly independent. For example, choosing the induced dipole and the two tensor-polarized mechanical densities, $\{p^{\mathrm{IMF}}_1,p^{\mathrm{IMF}}_2,p^{\mathrm{IMF}}_3\}$ and $\{s^{\mathrm{IMF}}_1,s^{\mathrm{IMF}}_2,s^{\mathrm{IMF}}_3\}$ as our minimal basis, the two scalar pressure and shear force densities can be determined as
\begin{align}
  p^{\mathrm{IMF}}_{(0,1)}(b) &=-\frac13 p^{\mathrm{IMF}}_2(b) -2p^{\mathrm{IMF}}_1(b) - \frac{1}{6m^2}\lap p^{\mathrm{IMF}}_3(b) \, , \label{p01From123}\\
  p^{\mathrm{IMF}}_{(0,0)}(b) &= -\frac13 p^{\mathrm{IMF}}_2(b) - 2p^{\mathrm{IMF}}_1(b) - \frac{1}{6m^2}\lap p^{\mathrm{IMF}}_3(b) -\frac{1}{m^2}\lap p^{\mathrm{IMF}}_1(b) \, , \label{p00From123} \\
  s^{\mathrm{IMF}}_{(0,1)}(b) &= -\frac13 s^{\mathrm{IMF}}_2(b) - 2s^{\mathrm{IMF}}_1(b) -\frac{1}{6m^2}\left(\lap-\frac{4}{b^2}\right)s^{\mathrm{IMF}}_3(b) \, , \label{s01From123}\\
  s^{\mathrm{IMF}}_{(0,0)}(b) &= -\frac13 s^{\mathrm{IMF}}_2(b) -2s^{\mathrm{IMF}}_1(b) - \frac{1}{6m^2}\left(\lap-\frac{4}{b^2}\right)s^{\mathrm{IMF}}_3(b) - \frac{1}{m^2}\left(\lap-\frac{4}{b^2}\right)s^{\mathrm{IMF}}_1(b) \, . \label{s00From123}
\end{align}
Thus, one can conclude from this that the two scalar IMF densities are not additional dynamical input as they can be reconstructed from the other three densities, given that the constraints in Eqs.~(\ref{2D_IMF_pressure_condition_1}--\ref{2D_IMF_shear_condition_2}) are satisfied. We chose the basis $\alpha=\{1,2,3\}$ here, but in principle, except for the $\alpha=\{ (0,0), (0,1), 1 \}$ basis, any other basis may also be used to construct entire 2D IMF densities in a relevant kinematic domain. This is due to the fact that the constraints in (\ref{2D_IMF_pressure_condition_1}--\ref{2D_IMF_shear_condition_1}) make this specific basis linearly dependent. In other words, at least one quadrupole-type pressure and shear force is necessary to construct 2D IMF densities.    

Eventually, the conversion from the reduced 2D BF distributions to the recoil-corrected ones can be achieved in coordinate space by applying 
\begin{align}
  p_n(b) &= \left[ 1 - \frac{\lap}{4 m^2}\right]^{-1/2} p_n^{\text{red}}(b) \, , \label{rec_red_p} \\
  s_n(b) &= \left[ 1 - \frac{1}{4 m^2}\bigg( \lap - \frac{4}{b^2} \bigg) \right]^{-1/2} s_n^{\text{red}}(b) \, . \label{rec_red_s}
\end{align}

Next, we want to obtain the entire 2D IMF mechanical densities from the 2D BF mechanical distributions. After converting the recoil-corrected 2D BF to the reduced ones by 
\begin{align}
  p_n^{\text{red}}(b) &= \left[ 1 - \frac{\lap}{4 m^2}\right]^{1/2} p_n(b) \, , \label{red_rec_p} \\
  s_n^{\text{red}}(b) &= \left[ 1 - \frac{1}{4 m^2}\bigg( \lap - \frac{4}{b^2} \bigg) \right]^{1/2} s_n(b) \, , \label{red_rec_s}
\end{align}
and rearranging Eqs.~(\ref{2D_BF_from_IMF_forces_1}--\ref{2D_BF_from_IMF_forces_3}) for IMF densities $\alpha=(0,1), 2, 3$, then inserting them into the constraint Eqs.~(\ref{2D_IMF_pressure_condition_2}--\ref{2D_IMF_shear_condition_2}), one obtains the following differential equations for $p^{\mathrm{IMF}}_1(b)$ and $s^{\mathrm{IMF}}_1(b)$

\begin{equation}
    \left[6-\frac{3}{2m^2}\nabla_\perp^2\right] p^{\mathrm{IMF}}_1(b) = 
    -\left[3 p^{\text{red}}_0(b) + p^{\text{red}}_2(b) + \frac{1}{2m^2}\nabla_\perp^2 p^{\text{red}}_3(b)\right] \, .
\end{equation}
\begin{equation}
    \left[6-\frac{3}{2m^2}\left(\nabla_\perp^2-\frac{4}{b^2}\right)\right]s^{\mathrm{IMF}}_1(b) = 
    -\left[3 s^{\text{red}}_0(b)+ s^{\text{red}}_2(b)+\frac{1}{2m^2}\left(\nabla_\perp^2-\frac{4}{b^2}\right) s^{\text{red}}_3(b)\right] \,.
\end{equation}
These equations reconstruct the Wigner rotation induced IMF dipole density and have the structure of an inhomogeneous modified Helmholtz equation, whose homogeneous solutions are the modified Bessel functions $I_n$ and $K_n$, with $n=0$ for $p_1^{\mathrm{IMF}}$ and $n=2$ for $s_1^{\mathrm{IMF}}$. The inhomogeneous equation can be solved by using the appropriate boundary conditions, which take into account the regular behavior of the densities in the limit $b\to 0$ and their decaying behavior at $b\to\infty$. Since the modified Bessel function $I_n$ grows exponentially as $b\to \infty$ and behaves regularly at $b=0$, while $K_n$ behaves singular at the origin and decays as $b\to\infty$, one obtains the solutions 
\begin{align}
    p^{\mathrm{IMF}}_1(b) &= -\frac{2m^2}{3} \int_0^\infty db'\, b'\, I_0(2mb_\mathrm{min}) K_0(2mb_\mathrm{max}) \left[3 p^{\text{red}}_0(b')+ p^{\text{red}}_2(b') + \frac{1}{2m^2}\nabla_\perp^2 p^{\text{red}}_3(b')\right]\, , \label{p1_from_2D_BF} \\
    s^{\mathrm{IMF}}_1(b) &= -\frac{2m^2}{3}\int_0^\infty db'\,b'\,I_2(2mb_\mathrm{min})K_2(2mb_\mathrm{max})\left[3 s^{\text{red}}_0(b')+ s^{\text{red}}_2(b')+\frac{1}{2m^2}\left(\nabla_\perp^2-\frac{4}{b'^2}\right) s^{\text{red}}_3(b')\right] \,, \label{s1_from_2D_BF}
\end{align}
where $b_\mathrm{min} = \text{min}(b, b')$ and $b_\mathrm{max} = \text{max}(b, b')$ are the minimum and maximum of two numbers, $b$ and $b'$, respectively. Once $p^{\mathrm{IMF}}_1(b)$ and $s^{\mathrm{IMF}}_1(b)$ are obtained, it is straightforward to obtain the rest of the 2D IMF densities by using relevant Eqs.~(\ref{2D_BF_from_IMF_forces_1}--\ref{2D_IMF_shear_condition_2}). We stress that, unlike obtaining the reduced 2D BF distributions from 2D IMF densities, which is a local operation, the inverse direction requires a nonlocal operation, an integral evaluation, as given in Eqs.~(\ref{p1_from_2D_BF}--\ref{s1_from_2D_BF}). 

\section{Forward and Inverse Abel transformations of forces}

Finally, we want to apply the Abel transformation to 3D BF distributions to obtain 2D BF distributions and the inverse Abel transformation to obtain 3D BF distributions from the 2D ones. Originally, for spin-1/2 particles, Abel tomography was used to relate 3D BF distributions to 2D IMF densities \cite{Panteleeva:2021iip}. For spin-1 particles, the Abel transformation is not sufficient to relate 3D BF distributions to the 2D IMF densities, as shown in \cite{Kim:2022wkc}. However, when combined with the results from the previous section, the Abel transformation and its inverse can be used to relate the 3D BF and the 2D IMF pressure and shear-force distributions through the reduced 2D BF distributions. This way, one finds an invertible map between the 3D BF distributions and 2D IMF densities in coordinate space. 

We proceed with writing the forward projection first. For each $n=0,2,3$, let us denote the three-dimensional and two-dimensional normal forces as
\begin{equation}
    N_n(r) = p_n(r)+\frac{2}{3} s_n(r) \, , \qquad N_n(b) = p_n(b) + \frac12 s_n(b) \, .
\end{equation}
Then, starting from the 3D BF distributions, one obtains the 2D BF distributions by applying them the Abel transformations \cite{Kim:2022wkc}
\begin{align}
  s_n(b) &= 2b^2\int_b^\infty\frac{dr}{r}\,\frac{s_n(r)}{\sqrt{r^2-b^2}} \,, \\
  N_n(b) &= \int_b^\infty dr\,\frac{r\,N_n(r)}{\sqrt{r^2-b^2}} \,. \label{Normal_Abel_transformation}
\end{align}
But, unlike the spin-1/2 case \cite{Panteleeva:2021iip}, it is not sufficient to produce 2D IMF densities from the 3D BF ones with a single Abel transformation. Once Abel transformation produces a 2D BF distribution, one first needs to remove the recoil correction effects with Eqs.~(\ref{red_rec_p}--\ref{red_rec_s}) to obtain the reduced 2D BF distributions, and then needs Eqs.~(\ref{2D_BF_from_IMF_forces_1}--\ref{2D_IMF_shear_condition_1}, \ref{p1_from_2D_BF}--\ref{s1_from_2D_BF}) to obtain individual 2D IMF densities. 

Since each 2D BF distribution is generated by a two-dimensional potential, $\widetilde{\DDbar}_n(b)$, it satisfies the two-dimensional equilibrium conditions for each $n$
\begin{equation}
  p'_n(b) + \frac12 s'_n(b) + \frac{1}{b} s_n(b)=0, \qquad N'_n(b)=-\frac{s_n(b)}{b} \,.
\end{equation}
Similarly, if one has 2D IMF densities at hand, or its minimal basis, one can recombine them as in Eqs.~(\ref{2D_BF_from_IMF_forces_1}--\ref{2D_BF_from_IMF_forces_3}) to produce the reduced 2D BF distributions, with which, after adding the recoil corrections in (\ref{rec_red_p}--\ref{rec_red_s}), one can obtain the 3D BF distributions via the inverse Abel transformations
\begin{align}
  s_n(r) &= -\frac{r^2}{\pi} \int_r^\infty\frac{db}{\sqrt{b^2-r^2}}\frac{d}{db}\left(\frac{s_n(b)}{b^2}\right) \,, \label{inverse_shear} \\
  p_n(r)+\frac23s_n(r) &= -\frac{2}{\pi}\int_r^\infty\frac{db\, N'_n(b)}{\sqrt{b^2-r^2}}
  = \frac{2}{\pi} \int_r^\infty\frac{db}{b}\,\frac{s_n(b)}{\sqrt{b^2-r^2}} \label{inverse_normal}\, .
\end{align}
This procedure is the direct spin-1 analog of the spin-1/2 Abel tomography relation \cite{Panteleeva:2021iip}. Once the Wigner spin rotation-induced mixing is undone, the three-dimensional normal and shear forces are determined solely by the corresponding 2D BF shear forces under the inverse Abel transformation. 

\section{Illustration}

In this section, we illustrate some of our findings using a quadrupole-type parametrization of the 3D BF form factors, a toy model also used in various works, for instance, in Refs.~\cite{Lorce:2018egm, Freese:2021czn, Kim:2022wkc}. In this model, we first want to show that the 2D IMF scalar densities $p^{\mathrm{IMF}}_\alpha$ and $s^{\mathrm{IMF}}_\alpha$ with $\alpha=(0,0), (0,1)$ can be obtained from the rest of the IMF densities. Then, we want to reconstruct the 3D BF pressure and shear forces from a minimal basis of 2D IMF densities, chosen here as $\alpha=1, 2, 3$.  

\begin{figure}[htbp]
    \centering
    \includegraphics[width=0.95\linewidth]{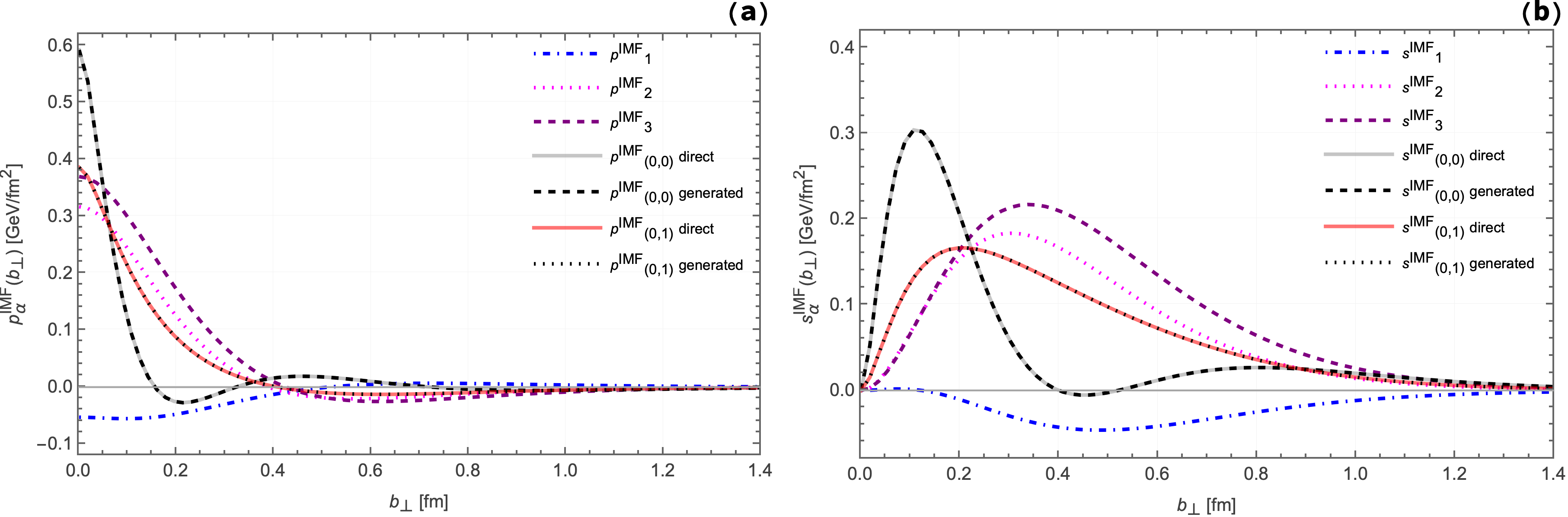}
    \caption{2D IMF (a) pressure and (b) shear force densities obtained directly from the Fourier transform of the IMF multipoles. The densities $p^{\mathrm{IMF}}_{(0,0)}, p^{\mathrm{IMF}}_{(0,1)}$ and $s^{\mathrm{IMF}}_{(0,0)}, s^{\mathrm{IMF}}_{(0,1)}$ are also obtained from Eqs.~(\ref{p01From123}--\ref{s00From123}) for comparison.}
    \label{fig1}
\end{figure}

To begin with, for simplicity, we take the same BF multipole form factors for $n=0,2,3$. 
\begin{equation}
  \DD_n(t)=\frac{\DD_n(0)}{(1-t/\Lambda^2)^4}, \quad \text{with} \quad \DD_0(0)=\DD_2(0)=\DD_3(0)=-1 \, , \label{multipole_parametrization}
\end{equation}
where $\Lambda$ is the cutoff value determined by the value $2m_\rho$, where $m_\rho$ is the mass of the rho meson. Using Eqs.~(\ref{IMF_multipoles}--\ref{GW}), one can obtain the full set of 2D IMF multipoles and we express the corresponding 2D IMF pressure and shear forces as
\begin{align}
  p^{\mathrm{IMF}}_\alpha(b)&=-\frac{1}{8m}\int_0^\infty\frac{d\Delta_\perp}{2\pi}\,\Delta_\perp^3 J_0(\Delta_\perp b)\,\DI_\alpha(-\Delta_\perp^2), \label{2D_IMF_pressure_Fourier} \\
  s^{\mathrm{IMF}}_\alpha(b)&=-\frac{1}{4m}\int_0^\infty\frac{d\Delta_\perp}{2\pi}\,\Delta_\perp^3 J_2(\Delta_\perp b)\,\DI_\alpha(-\Delta_\perp^2), \, \label{2D_IMF_shear_Fourier}
\end{align}
with $\alpha=(0,0),(0,1),1,2,3$. However, those five IMF multipoles are not independently parametrized, as they satisfy the constraints in Eqs.~(\ref{2D_FFconstraints_1}--\ref{2D_FFconstraints_2}). Noting that 
\begin{align}
  \nabla_\perp^2 p^{\mathrm{IMF}}_\alpha(b)&=\frac{1}{8m}\int_0^\infty\frac{d\Delta_\perp}{2\pi}\,\Delta_\perp^5 J_0(\Delta_\perp b)\,\DI_\alpha(-\Delta_\perp^2),\\
  \bigg(  \nabla_\perp^2 - \frac{4}{b^2}\bigg) s^{\mathrm{IMF}}_\alpha(b)&=\frac{1}{4m}\int_0^\infty\frac{d\Delta_\perp}{2\pi}\,\Delta_\perp^5 J_2(\Delta_\perp b)\,\DI_\alpha(-\Delta_\perp^2),
\end{align}
one can easily derive scalar pressure and shear forces from Eqs.~(\ref{p01From123}--\ref{p00From123}) and Eqs.~(\ref{s01From123}--\ref{s00From123}), respectively, from the basis $\{p^{\mathrm{IMF}}_1, p^{\mathrm{IMF}}_2, p^{\mathrm{IMF}}_3\}$ and $\{s^{\mathrm{IMF}}_1, s^{\mathrm{IMF}}_2, s^{\mathrm{IMF}}_3\}$. In Fig.~(\ref{fig1}), we prepare the same plot as in \cite{Kim:2022wkc} to illustrate the entire set of 2D IMF mechanical densities, directly obtained from the IMF multipole form factors, using Eqs.~(\ref{2D_IMF_pressure_Fourier}--\ref{2D_IMF_shear_Fourier}). In addition to them, we also add the curves $p^{\mathrm{IMF}}_{(0,0)}, p^{\mathrm{IMF}}_{(0,1)}$ and $s^{\mathrm{IMF}}_{(0,0)}, s^{\mathrm{IMF}}_{(0,1)}$ obtained from Eqs.~(\ref{p01From123}--\ref{s00From123}) by using the chosen basis, for comparison. We find that the results are numerically identical, confirming that those scalar densities are linearly dependent on the chosen basis. 

\begin{figure}[htbp]
    \centering
    \includegraphics[width=0.95\linewidth]{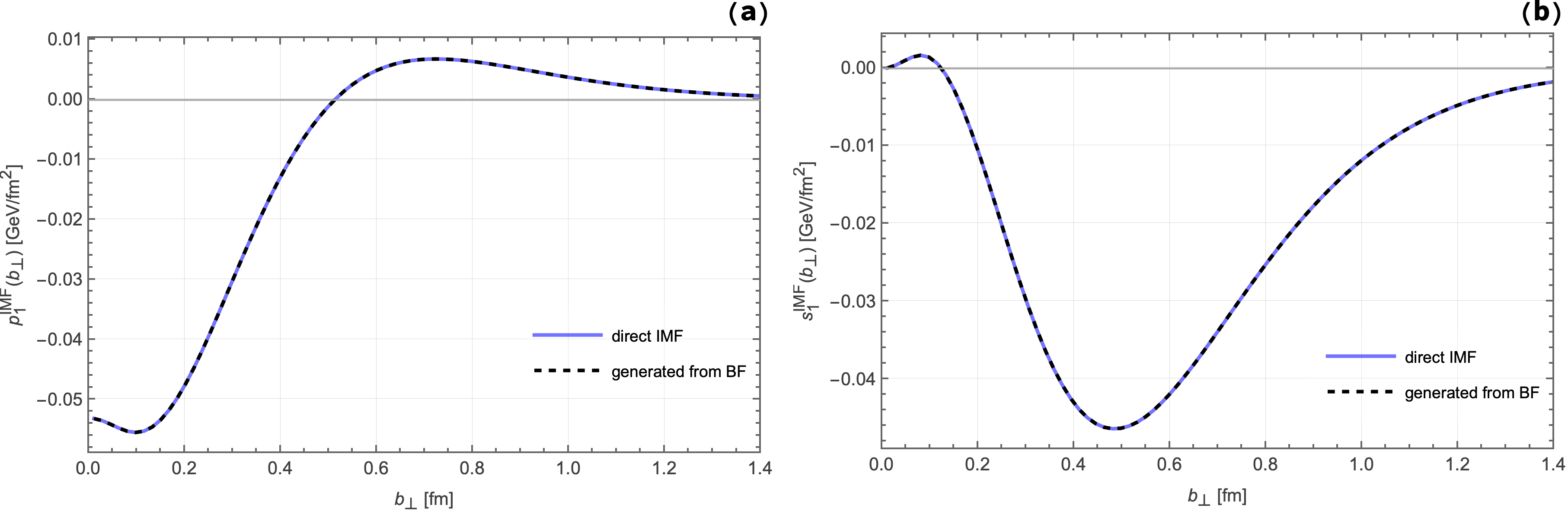}
    \caption{Comparison of the 2D relativistically induced IMF dipole densities (a) pressure $p^{\mathrm{IMF}}_1(b)$ and (b) shear force $s^{\mathrm{IMF}}_1(b)$, obtained first from IMF multipole by Fourier transformation (direct IMF), second reconstructed from BF distributions by solving the Helmholtz equations (generated from BF).}
    \label{fig2}
\end{figure}

Next, we compare the relativistically induced IMF dipole pressure and shear forces $p^{\mathrm{IMF}}_1(b)$ and $s^{\mathrm{IMF}}_1(b)$ by using two methods: first, we directly calculate them using Eqs.~(\ref{2D_IMF_pressure_Fourier}--\ref{2D_IMF_shear_Fourier}); second, we use the integral expressions in Eqs.~(\ref{p1_from_2D_BF}--\ref{s1_from_2D_BF}) obtained by solving the associated Helmholtz equations. Fig.~(\ref{fig2}) shows that both methods return identical results, confirming that 3D Breit frame distributions encode the relativistically induced IMF dipole densities.   

\begin{figure}[htbp]
    \centering
    \includegraphics[width=0.95\linewidth]{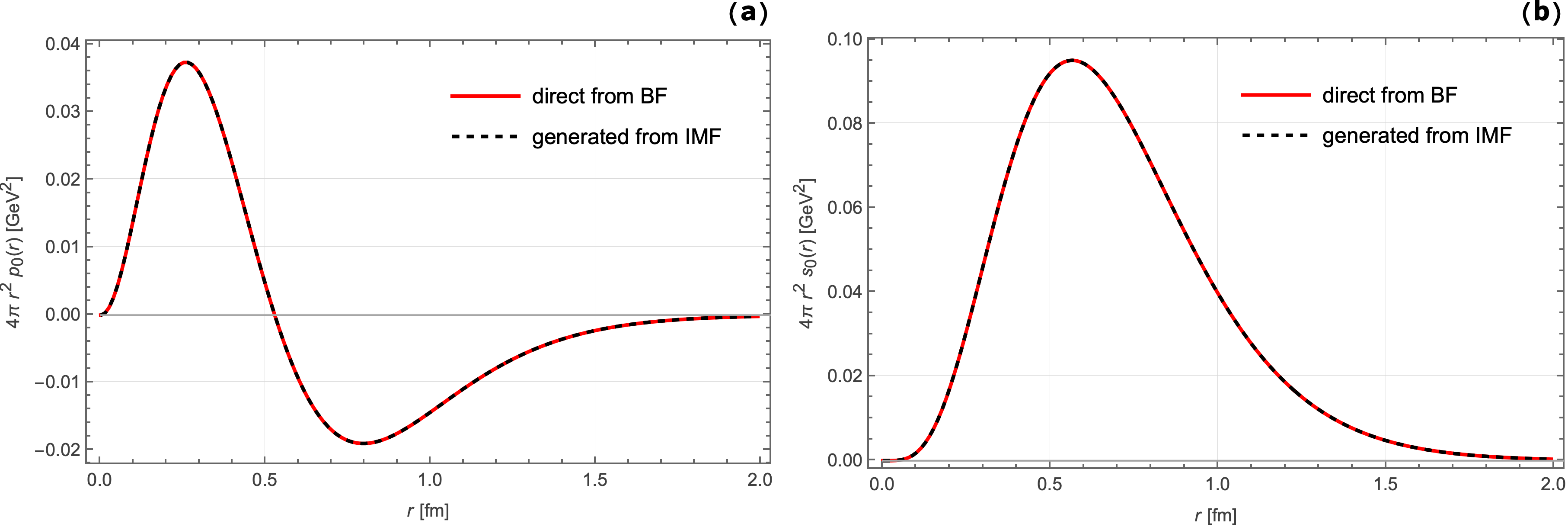}
    \caption{3D BF (a) pressure and (b) shear force distributions obtained directly from the Fourier transform of the BF multipoles and compared with the ones generated from 2D IMF densities.}
    \label{fig3}
\end{figure}

Last, we obtain the 3D BF pressure and shear force distributions from the basis of 2D IMF densities $\{p^{\mathrm{IMF}}_1, p^{\mathrm{IMF}}_2, p^{\mathrm{IMF}}_3\}$ and $\{s^{\mathrm{IMF}}_1, s^{\mathrm{IMF}}_2, s^{\mathrm{IMF}}_3\}$ by first recombining them to obtain the reduced 2D BF distributions, then incorporating the recoil factor, and then applying the inverse Abel transformations. We compare these results with the direct Fourier transform of the BF multipole form factors using (\ref{multipole_parametrization}).
The 3D BF distributions from the multipole form factors can be written as
\begin{align}
    p_n(r)&=-\frac{1}{6m}\int_0^\infty \frac{d\Delta}{2\pi^2}\,\Delta^4 j_0(\Delta r)\frac{m}{E}\DD_n(-\Delta^2) \, , \label{3D_pressure_direct} \\
    s_n(r)&=-\frac{1}{4m}\int_0^\infty \frac{d\Delta}{2\pi^2}\,\Delta^4 j_2(\Delta r)\frac{m}{E}\DD_n(-\Delta^2) \,. \label{3D_shear_direct}
\end{align}
On the other hand, the 3D BF distributions can also be obtained by recombining 2D IMF densities into the reduced 2D BF ones, as in Eqs.~(\ref{2D_BF_from_IMF_forces_1}--\ref{2D_BF_from_IMF_forces_3}), adding recoil corrections, and taking the inverse Abel transformation in Eqs.~(\ref{inverse_shear}--\ref{inverse_normal}).
Obtaining the recoil corrected 2D BF mechanical distributions from the reduced ones can be achieved by applying the Hankel transforms 
\begin{align}
    \hat{p}^{\text{red}}_n (\Delta_\perp) &= 2\pi\int_{0}^\infty db' \, b' \, J_0(\Delta_\perp b') \, p_n^{\text{red}}(b') \, , \\
    \hat{s}^{\text{red}}_n (\Delta_\perp) &= 2\pi\int_{0}^\infty db' \, b' \, J_2(\Delta_\perp b') \, s_n^{\text{red}}(b') \, ,
\end{align}
where the orders of the Hankel transformations are determined by the transformation properties of pressure and shear forces under rotations on the transverse plane. Taking the inverse Hankel transforms by including the recoil corrections then yields
\begin{align}
    p_n (b) &= \int_{0}^\infty \frac{d\Delta_\perp}{2\pi} \,\Delta_\perp \, J_0(\Delta_\perp b) \, \bigg( 1+\frac{\Delta_\perp^2}{4m^2}\bigg)^{-\frac{1}{2}} \hat{p}_n^{\text{red}}(\Delta_\perp) \, , \\
    s_n (b) &= \int_{0}^\infty \frac{d\Delta_\perp}{2\pi}\, \Delta_\perp \, J_2(\Delta_\perp b) \, \bigg( 1+\frac{\Delta_\perp^2}{4m^2}\bigg)^{-\frac{1}{2}} \hat{s}_n^{\text{red}}(\Delta_\perp) \, .
\end{align}
Fig.(\ref{fig3}) illustrates the 3D pressure and shear force distributions, weighted by $4\pi r^2$, obtained from two different methods. First, we compute it directly from Eqs.~(\ref{3D_pressure_direct}--\ref{3D_shear_direct}). Second, we use the recombined IMF densities by using the minimal basis, add recoil corrections, and perform the inverse Abel transformation, as given in Eqs.~(\ref{inverse_shear}--\ref{inverse_normal}). Both methods yield the same results, confirming that the chosen minimal basis of 2D IMF densities is sufficient to reconstruct the 3D BF pressure and shear force distributions.  

\section{Conclusions}
In this work, we analyzed the relationship between pressure and shear forces in the 3D Breit frame and the 2D infinite-momentum frame. As explicitly shown in Ref.~\cite{Kim:2022wkc}, at the multipole form factor level, there is a mismatch in the number of form factors obtained between the two frames due to Wigner rotation mixings on the IMF side. The 3D BF multipole form factors have three, whereas the IMF ones have five. Since the IMF multipole form factors can be fully expressed in terms of 3D BF form factors \cite{Kim:2022wkc}, the apparent mismatch is due to two constraints on the IMF side that must be considered. These relations at the form factor level propagate into the pressure and shear force distributions. We argued that the relation between the two frames in coordinate space becomes clear after recombining or separating Wigner spin-rotation effects on the IMF side in order to relate them to the reduced 2D BF distributions. Once this step is done, one can apply the forward and inverse Abel transformations to relate 3D BF distributions and 2D IMF densities through the 2D BF distributions. The absence of such mixings and recoil corrections in spin-1/2 hadrons allows the connection between 3D BF and 2D IMF to be made via a single Abel transformation \cite{Panteleeva:2021iip}. For spin-1 hadrons, we showed that obtaining reduced 2D BF distributions from the IMF ones is particularly straightforward, as it is a local operation that recombines IMF densities at the same transverse point. Once one obtains the recoil-corrected 2D BF distributions, applying the inverse Abel transform then yields the 3D BF distributions. On the other hand, after applying the Abel transformation to the 3D BF distributions, one obtains the 2D BF distributions. Reconstructing the entire 2D IMF densities from the 2D BF distributions requires removing the recoil correction terms and solving a modified inhomogeneous Helmholtz equation. The solution to this equation encodes the relativistically induced effects present in the $p^{\mathrm{IMF}}_1$ and $s^{\mathrm{IMF}}_1$ densities with which one can reconstruct the other IMF densities. Finally, we supported our results with numerical evidence by studying a quadrupole-type parametrization of form factors. We showed that choosing a three-element basis on the IMF side, here namely $\{ p^{\mathrm{IMF}}_1, p^{\mathrm{IMF}}_2, p^{\mathrm{IMF}}_3\}$ and $\{ s^{\mathrm{IMF}}_1, s^{\mathrm{IMF}}_2, s^{\mathrm{IMF}}_3\}$, reproduced the scalar-type densities due to the constraint relations among them. We also showed that the specific IMF basis chosen here can reproduce 3D BF distributions via the inverse Abel transformation. In principle, any other three-element basis can be chosen to reconstruct the entire IMF densities, except the basis $\alpha = \{ (0,0), (0,1), 1 \}$, which lacks a tensor-polarized density.

\ \\
{\bf Acknowledgments.}
I am grateful to Adam Freese, Simonetta Liuti, Cédric Lorcé, and Frank Vera for helpful discussions. This work was supported by the EXCLAIM collaboration through DOE grants DE-SC0016286 and DE-SC0024644.

\newpage


\end{document}